\documentclass{article}

\usepackage[utf8]{inputenc}
\usepackage{amsmath}
\usepackage{authblk}
\usepackage{lineno}

\usepackage[sorting=none]{biblatex}
\bibliography{references}

\usepackage{graphicx}
\usepackage{csvsimple}
\usepackage{longtable}
\usepackage[hypertexnames=false]{hyperref}

\begin{document}

\providecommand{\keywords}[1]
{
  \small	
  \textbf{\textit{Keywords---}} #1
}

\title{Winterization of Texan power system infrastructure is profitable but risky}

\author[1]{Katharina Gruber}
\author[2]{Tobias Gauster}
\author[1]{Peter Regner}
\author[2]{Gregor Laaha}
\author[1]{Johannes Schmidt}
\affil[1]{Institute for Sustainable Economic Development, University of Natural Resources and Life Sciences, Vienna, Austria}
\affil[2]{Institute of Statistics, University of Natural Resources and Life Sciences, Vienna, Austria}

\renewcommand\Affilfont{\itshape\small}

\date{April 2021}

\maketitle


\begin{abstract}

We deliver the first analysis of the 2021 cold spell in Texas which combines temperature dependent load estimates with temperature dependent estimates of power plant outages to understand the frequency of loss of load events, using a 71 year long time series of climate data. The expected avoided loss from full winterization is 11.74bn\$ over a 30 years investment period. We find that large-scale winterization, in particular of gas infrastructure and gas power plants, would be profitable, as related costs for winterization are substantially lower. At the same moment, the necessary investments involve risk due to the low-frequency of events – the 2021 event was the largest and we observe only 8 other similar ones in the set of 71 simulated years. Regulatory measures may therefore be necessary to enforce winterization.

\end{abstract}

\keywords{
Texas, extreme event, power systems, winterization}
\medskip


Weather extremes such as storms can significantly affect the reliability of power systems \cite{Bennett2021}. The increasing use of variable renewable energies additionally exposes power systems to hazards caused by weather extremes \cite{Thornton2017,HOLTINGER2019695}. However, recently it was a gas power dominated system which was deeply impaired by a weather extreme: a cold spell over Texas, between February 10\textsuperscript{th} and February 20\textsuperscript{th}, 2021 with temperatures far below 0° caused a failure of large parts of the Texan power system. The combination of extraordinarily high winter electricity demand and more importantly the failure of significant power generation capacities, both due to low temperatures, resulted in up to 4.5 millions of Texans being cut-off from their electricity supply \cite{Pallone2021}.

Wu et al. \cite{wu2021opensource} provide an open source grid simulation to conduct a very detailed analysis of the 2021 event, but do not put the 2021 event into a long-term climatic context. In contrast, Doss-Gollin et al.  \cite{DossGollin2021} have shown that lower temperatures than in February 2021 have been observed in the past 71 years, and heating demand predicted from temperature data would also have been higher in the past, although the 2021 frost event was comparably long.
There is therefore a striking gap between the occurrence probability of such an event, its large scale economic and social cost, and the lack in winterization efforts. Hence, we assess here how avoided loss due to winterization compares to its cost. We do so in a simulation framework which allows to estimate the probability distribution of loss of load events, thus being able to derive the uncertainty of the magnitude of avoided loss within the investment period. Technically, we combine estimates of temperature dependent load with a model of power plant outages, taking into account 71 years (1950--2021) of past climate from reanalysis data. Climate change, of course, may have an impact on temperatures. We therefore also assess if trends in the occurrence of extremely cold temperatures and loss of load events can be observed. Furthermore we conduct an extensive sensitivity analysis to show uncertainties arising from our modelling choices.

\section*{What happened in February 2021?}
\begin{figure}[!ht]
\centering
\includegraphics[scale=0.41,trim={0.29cm 0.30cm 0.25cm 0.25cm},clip]{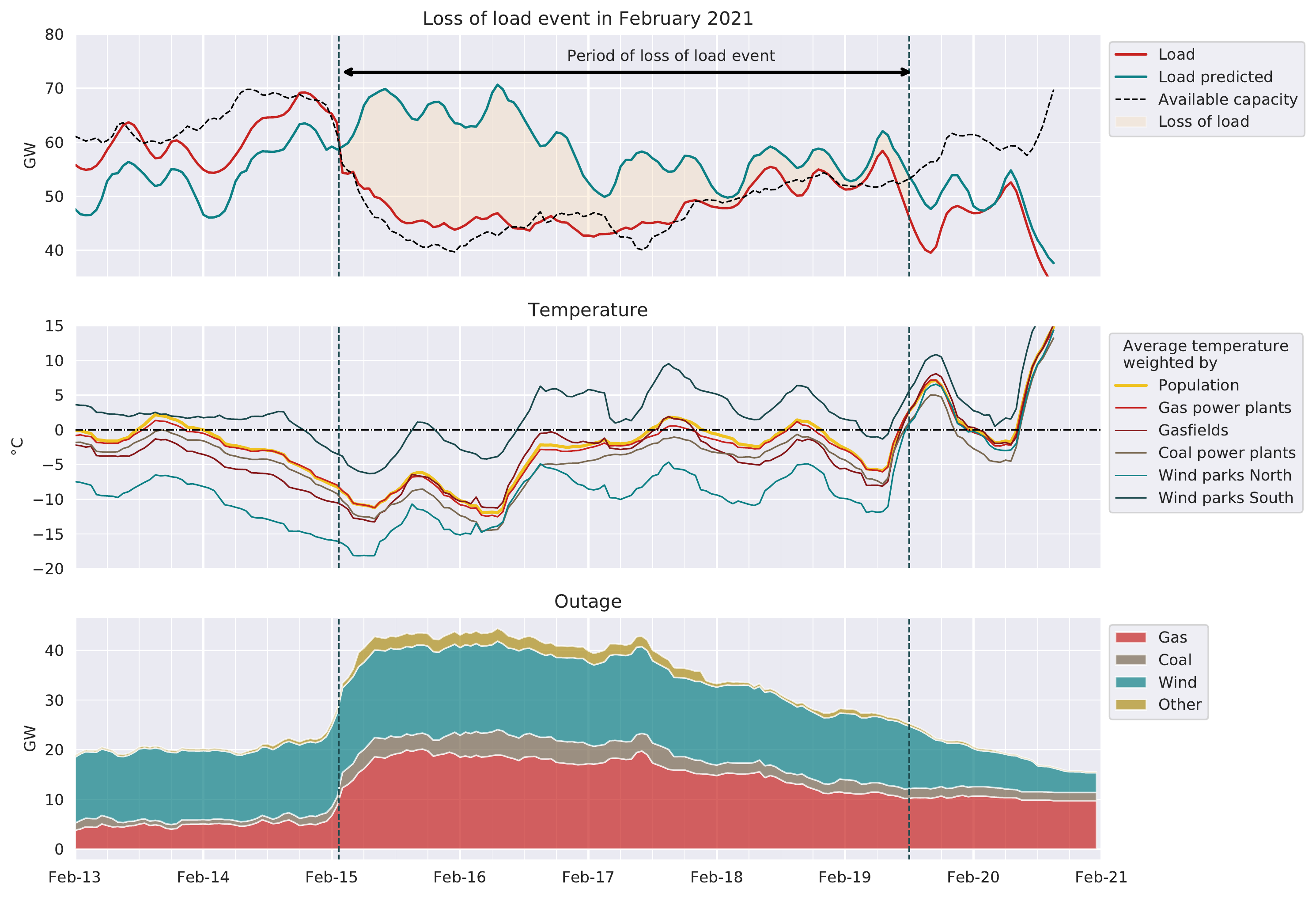}
\caption{Observed load, predicted load, available capacity, temperatures and outages during the February 2021 event (based on ERCOT load and outages \cite{ERCOTload,outages}, ERA5 temperatures \cite{ERA5}, and our load prediction model combining the two)}
\label{fig:load_outage_temp}
\end{figure}

Starting on 10\textsuperscript{th} of February 2021, temperatures in Texas began to fall, causing load to increase from around 40 GW to over 70 GW by February 14\textsuperscript{th}--15\textsuperscript{th}. On February 15\textsuperscript{th} the aggravating frost reached a critical level where substantial shares of generation capacities began to fail. Available capacities dropped below demand leading to a sustained power generation capacity deficit (Figure \ref{fig:load_outage_temp}). Consequently, rolling blackouts had to be implemented to stabilize the grid and prices at the power market increased to the upper limit of 9000\$/MWh. The deficit event continued until 20\textsuperscript{th} February when rising temperatures allowed the system to recover.

The highest load forecast in the February 2021 event was well above the highest load observed in winter in the period 2004--2020\footnote{The values of loss of load and load prediction in this section rely on our simulation and may therefore differ from ERCOT reports to some extent. Served load and plant outages are taken from ERCOT. As we focus on estimating the long-term frequency of such events, we did not aim at reproducing the February 2021 model in highest detail.}. Our estimate is in the range of observed extreme summer loads (see Appendix Figure \ref{fig:load_pred_obs_vs_Feb2021}). In contrast, ERCOT forecasts of peak load during the cold spell have been higher by about 4GW compared to our estimates \cite{IEA2021}, indicating an almost record-high predicted load on the network. Our estimates therefore have to be considered to be conservative.

Besides leading to high electricity load, the low temperatures also caused substantial outages of generation capacities. Gas capacity failures were responsible for the largest share in power outages. Out of 62GW of thermal capacity expected to be available in winter by ERCOT\footnote{Actual capacity available before the event has been higher, but transient stability requirements and reactive power demands have reduced the amount of load that has been covered in the system \cite{wu2021opensource}}, 25.4GW of thermal capacity failed in total, with the share of gas being 20.1GW. Based on the predicted demand and the observed load, we estimate that in total 1.19TWh of load were affected by blackouts. The maximum total outage capacity, including wind power, was 44.4GW, causing 24GW of peak lost load, when using our load prediction model. Loss of load occurred in 107 hours in the period from 2021-02-15 02:00 to 2021-02-19 12:00 local time.

Outages of gas generation capacities started to increase rapidly at a gas power plant weighted temperature of -8.8°C, which is a record low temperature compared to the past 17 years (see Figure \ref{fig:temp_gasfields_powerplants}). The outages were not only related to freezing of power plants, but also of gas supply infrastructure, including gas production equipment at gas fields. Power plants started failing rapidly when temperature weighted by gas fields dropped below -10.9°C. In the period 2004--2021, when no other outage events comparable to the one in 2021 was observed, this is a record low gas field weighted temperature (see Figure \ref{fig:temp_gasfields_powerplants}). Therefore, gas supply infrastructure may have played an important role in the outage events. This is confirmed by ERCOT, which classified around 8GW of outages being related to limited fuel supply \cite{OutageCauses}.

Coal generation capacity came offline at average temperatures weighted by coal plant locations of below -10.2°C. This temperature is at the very lower end of the temperature distribution in the period 2004--2020. For both technologies, coal and gas, recovery time was substantial. Even when temperatures recovered back to over 0°C, 11.3GW of thermal power plants, i.e. 18\% of total available thermal capacity, stayed offline for another 16 hours.

Temperatures weighted by wind power plant locations indicate that the failure of wind power plants may be a more frequent event. While temperatures at wind parks in Southern Texas were at the very lower end of the temperature range observed in the period 2004--2020, the average wind park temperature in Northern Texas was just below 0°C and well within the range of previously observed low temperatures. Compared to thermal power generation, wind power capacities began to fail much earlier and at higher temperatures. On February 13\textsuperscript{th}, when gas outages summed up to only 5GW, ERCOT already reports 13GW of wind power outages (Figure \ref{fig:load_outage_temp}).

\section*{How extreme was the February 2021 event?}
\begin{figure}[!ht]
\centering
\includegraphics{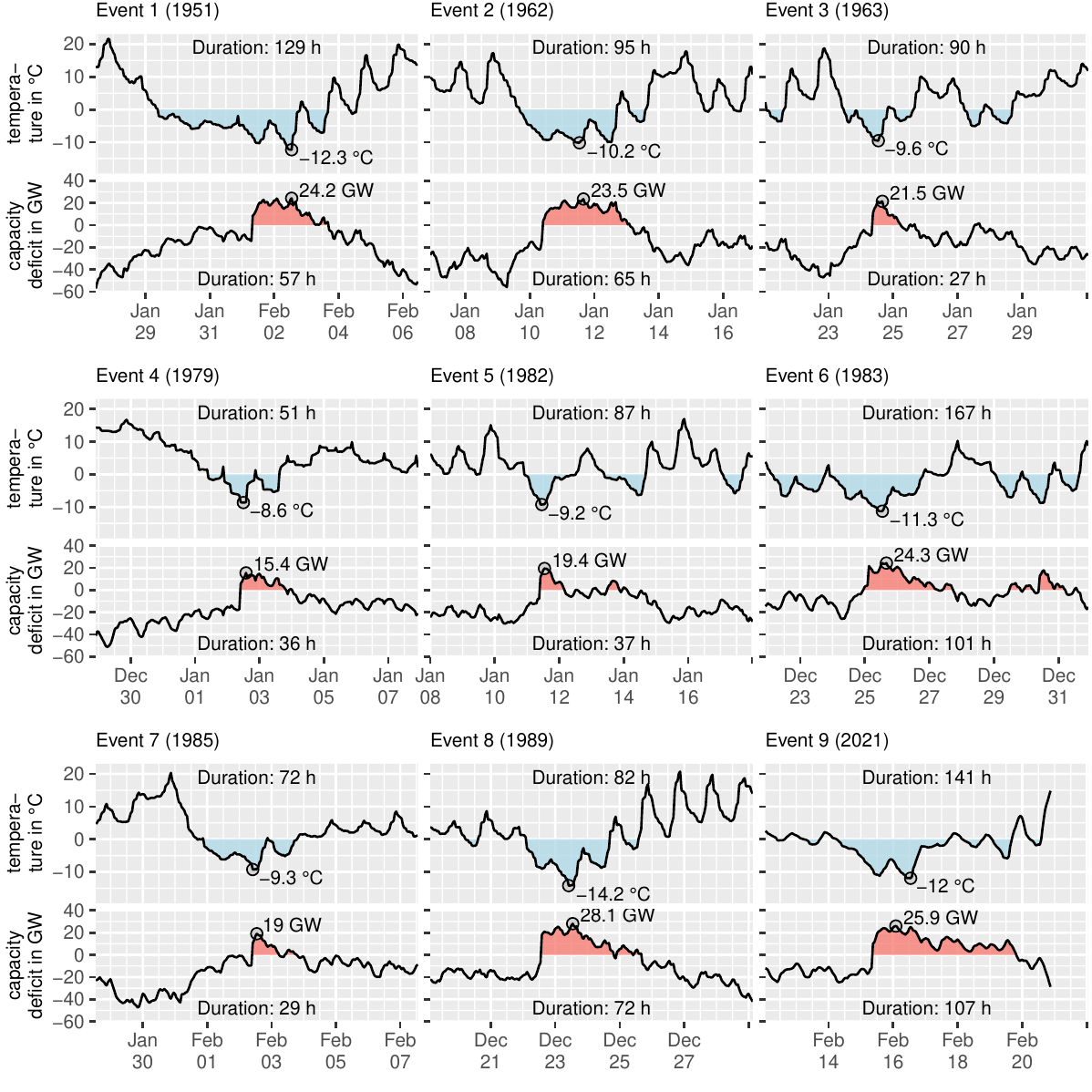}
\caption{Population weighted temperature and predicted capacity deficit of severe frost events within the 1950-2021 period. Labels in the graph refer to temperature minima and deficit maxima}
\label{fig:deficit_events}
\end{figure}

Our simulations of loss of load events using climate data from 71 years shows that the 2021 event was a record one\footnote{Please observe that the results on the 2021 event in this section differ slightly from the previous section, as we used simulated plant outages instead of outages provided by ERCOT here.}. In total, we estimate that eight other severe power deficit events would have occurred in the current system assuming climate from the period 1950-2021 (see Figure \ref{fig:deficit_events}). The second largest power deficit event at 0.98 TWh is predicted when using climate data from 1989. 

In our model predictions, the loss of load event has a duration of 107 hours, and causes an aggregated deficit of 1.39TWh, at a peak capacity deficit of 25.9GW. There are several events with similar peak capacity deficits identified in the 1950--2001 period, but none of the events has a comparably long duration and a comparably high amount of loss of load (Figure \ref{fig:deficit_events}). 1989 was the last time a similar frost event occurred. This long break in frost events of a significant magnitude may explain why the recent event hit an insufficiently winterized power production system. 

The 2021 record high loss of load is not caused by the frost magnitude alone but by a combination of a long, relatively cold frost event and an inopportune timing of the frost peak. According to Figure \ref{fig:load_outage_temp} the system failure occurred early and was prolonged by a long frost period afterwards. This is in contrast to other years when temperatures recovered more quickly after temperature minima had been reached (e.g. in 1951 and 1963). 
This finding is supported by the extreme value statistics of the frost spells shown in Figure \ref{fig:extreme_statistics}. 2021 was the longest frost event in seven decades. It has a return period of 141 years. Other events, however, were colder (1951, 1989) or had higher frost sums (1951). 

In terms of load, the highest predicted winter load in 2021 was slightly higher than the highest predicted winter load in the complete 71 year time series (Figure \ref{fig:load_pred_obs_vs_Feb2021}), although temperatures were lower in the 1989 event. A particular combination of time of day, day of week, and low temperature caused this particularly high load in 2021.

It may also have been expected that frost events would have decreased due to global warming. However, our analysis does not show any significant trend in the loss of load time series (Figure \ref{fig:trend-9events}). Still, average temperatures in Texas significantly increased due to climate change since 1951 (Figure \ref{fig:avg_temp_trend}). This result is confirmed by others, however, the increase in mean temperature is not genuinely transferable to extreme temperatures \cite{extremeTrend}. A stratified analysis of annual frost events (minimum annual temperature) below temperature thresholds from 0 to -10 °C reveals that there is indeed no significant observed change of severe frost events below -2 °C (Section \ref{section:ext_temp_trends}). Only very mild frost events showed a significant attenuation (2.6 °C over the past seven decades), but such events are irrelevant for frost related failures of the power system comparable to the 2021 event. 

Overall, extreme value statistics show that the event of 2021 was severe because of its long frost duration and its frost dynamics. Against the background of seven decades of observed climate data, the frost event had to be expected, especially as we could not find evidence for a decrease of severe frost events due to climate change. This suggests that similar events comparable to the one in 2021 have to be expected again and need to be mitigated even in a future, warmer climate.

\section*{Comparing avoided loss to costs of winterization}
A bootstrap of loss of load events from our 71 years of simulated capacity deficits yields an expected loss of load due to a cold weather event in a typical 30 year investment period of 2.55TWh with a 68\% confidence interval of [1.09TWh, 4.02TWh]\footnote{This confidence interval is a result of bootstrapping different loss of load events from different weather years and does not take into account other model uncertainties.}. The current maximum market price regulated by ERCOT is 9\,000\$/MWh \cite{ERCOTprice9000}. The avoided loss for electricity consumers under full winterization is therefore 11.74bn\$ at a 5\% discount rate\footnote{We emphasize here that this loss does not represent full societal cost of load shedding. However, the value is an indicator of incentives to winterize in the system.}.

\begin{figure}[!ht]
\centering
\includegraphics[scale=0.70,trim={0.1cm 0.1cm 0.1cm 0.1cm},clip]{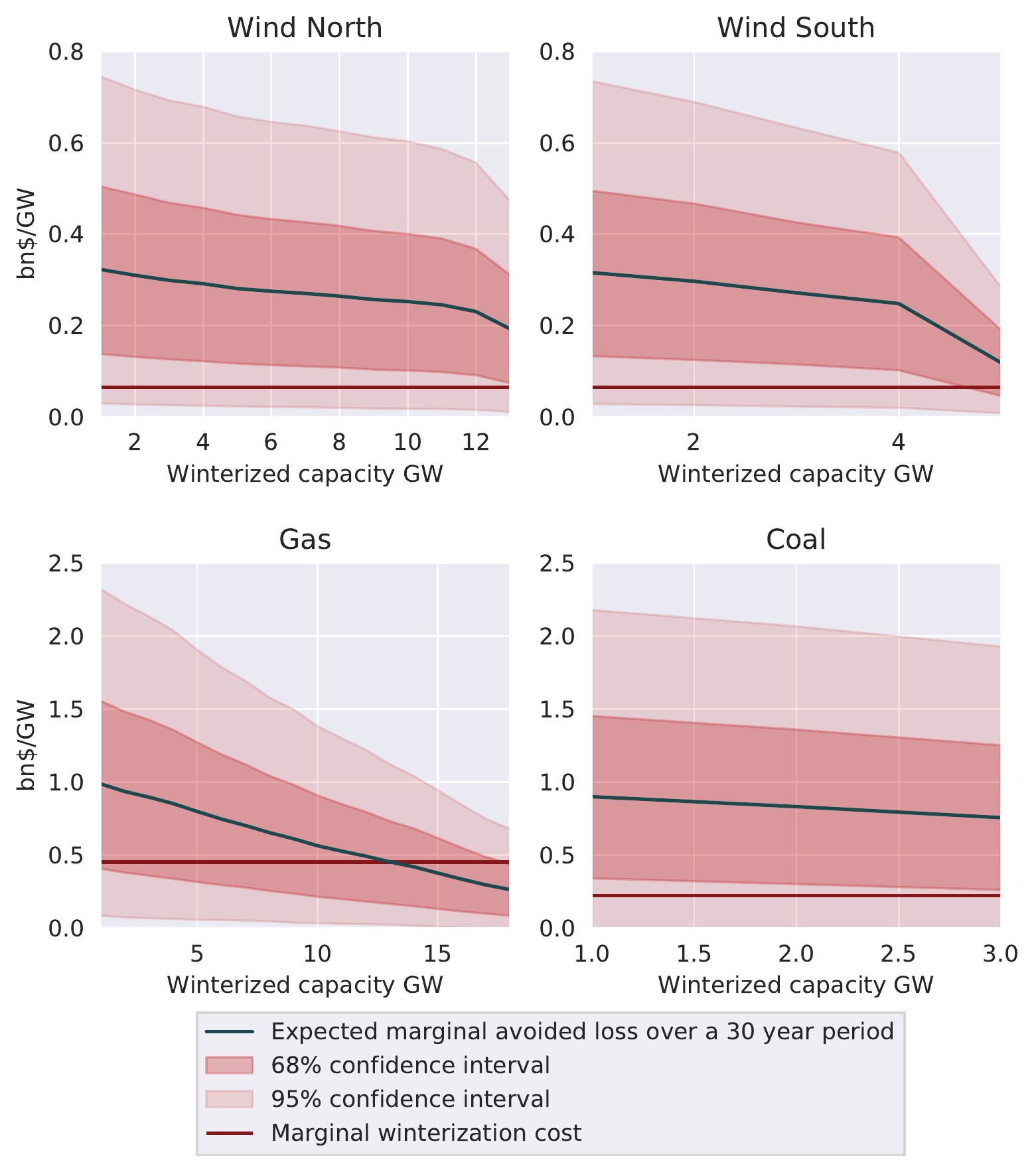}
\caption{Comparison of marginal avoided loss (bn\$/GW of winterized capacity in a 30 year period) to marginal costs for winterizing existing power generation capacity}
\label{fig:marginal_winterize}
\end{figure}

Expected marginal avoided loss is higher than winterization cost for most of the failed gas, wind, and coal power infrastructure (Figure \ref{fig:marginal_winterize}). For the first winterized GW of gas power capacity, expected marginal avoided loss over a 30 years period is at 0.98bn\$/GW, but drops to 0.52bn\$/GW at 11GW of winterization. Marginal avoided loss for coal power plant winterization is slightly lower per GW, and significantly lower for wind power. For all technologies, the spread of the marginal avoided loss is high: The 68\% confidence interval is at double or half of the expected marginal avoided loss. In 1.7\% of all cases, there is no deficit event in a 30 year period, which is the worst case scenario in case of investment in winterization, because there is no avoided loss.
 
Significant winterization measures can be implemented under our estimates of expected marginal avoided loss. In particular, we estimate that the winterization of gas wells in combination with winterization of gas power plants will cost about 450M\$/GW (see section \ref{section:winterization_costs}).
This cost is below the marginal avoided loss up to the 13\textsuperscript{th} GW of winterized capacity. Winterization of coal and wind power plants is significantly cheaper, as fuel supply infrastructure does not have to be winterized. Winterization costs assumed at 10\% of initial plant investments of coal power plants are far below marginal avoided loss up to full winterization of all failed coal capacity. In fact, one could even assume winterization cost of 30\% of initial plant investments and winterization cost would still be lower than the marginal avoided loss for the completely winterized capacity. For wind turbines, our estimates of marginal avoided loss are half those of coal, but are still substantially higher than the costs of winterization which are reported to be 5\% of investment costs \cite{wind_winterize}.

\section*{How reliable are our estimates?}
\begin{figure}[!ht]
\centering
\includegraphics[scale=0.61,trim={0cm 0cm 0cm 0cm},clip]{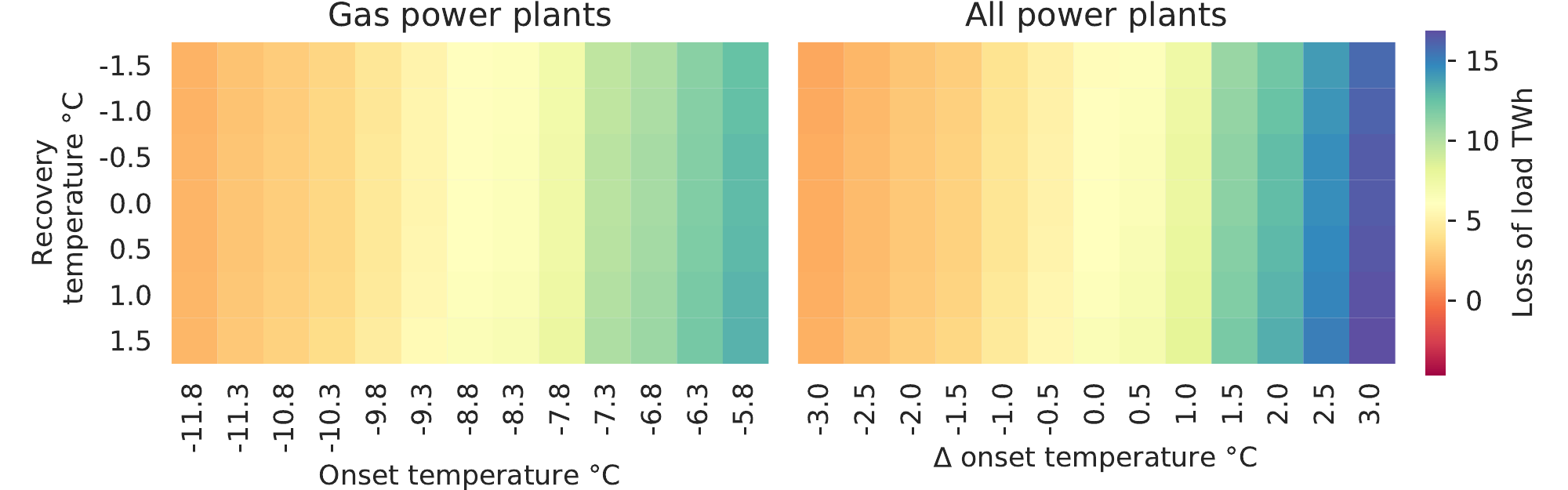}
\caption{Sensitivity of the accumulated loss of load between 1950 and 2021 by generation type to onset temperature and recovery temperature used in the outage model for a variation of gas threshold temperatures (left) or all threshold temperatures together (right)}
\label{fig:sensitivity_temp_gas_all}
\end{figure}

Estimates of loss of load events depend mainly on the assumptions on outage temperatures for gas power plants, as their failure has by far the biggest impact on the magnitude of loss of load events. An increase in outage temperatures from -8.8°C to -5.8°C  increases loss of load from 6.04--6.48TWh to 12.61--13.18TWh in 71 years, depending on the respective recovery temperature  (see Figure \ref{fig:sensitivity_temp_gas_all}). In contrast, the outage temperatures assumed for wind and coal power plants have a minor impact on the estimates of loss of load, if they are changed for one technology only (in the range of 5.67--6.34TWh). A concurrent increase of outage temperatures of all power generation technologies changes loss of load from 5.92--6.54TWh to 15.92--16.94TWh when increasing the outage temperature by 3°C. In contrast, recovery temperatures do not change our estimates of loss of load significantly, as loss of load shows less than 1TWh difference within a variation of outage temperature of 3°C. Lowering outage temperatures decreases our estimates likewise. For a temperature decrease of 3°C, the loss of load predictions are less than a third of our base estimate (1.61--1.90TWh).

Our finding that 2021 was the largest simulated loss of load event in the period 1950--2021 is sensitive to the assumed outage temperature. If outage temperatures of power plants are simultaneously increased by 1.5°C, 2021 becomes the second largest event while 1983 becomes the largest event. If additionally the recovery temperature falls by 1.5°C, 2021 becomes the third largest event, as 1983 and 1989 are larger. The loss of load event in 2021, however, in none of the sensitivity simulations has a rank lower than 3 in terms of total loss of load.

While it is certain that climate change affects average temperatures, our analysis indicates that it may not have caused a trend of decreasing extreme cold events. Nevertheless, we assessed how the number and magnitude of power deficit events would change if temperature extremes would follow the same trend as average temperature. Under such assumptions, we find that the number of severe loss of load events is reduced from 9 to 8, while the loss of load drops from 6.11TWh to 4.37TWh and 3.66TWh, assuming 2021 and 2050 as years for simulation, respectively (see section \ref{section:ext_temp_trends} for modeling details). We emphasize here that our approach is in contrast to observations (see \ref{section:trends_extreme}) and should only be understood as sensitivity analysis. Nevertheless, even under such a scenario, the loss of load is still significant and - a however reduced - extent of winterization would be profitable. 

Instead of using aggregated outage curves by technology, as we do in our analysis, we have also developed plant level outage curves. When using those, the number of events increases from 9 to 48, however, the total deficit increases from 6.11TWh to 9.72TWh only, as most events are minor. Such a high number of outage events is unrealistic, as they were not observed in the period 2004--2021, for which data is available. We conclude that at the moment the used outage data provided by ERCOT does not allow to derive temperature dependent plant level outages that represent reasonable loss of load events. Our aggregated approach also is associated with significant uncertainty (see \ref{subsection:outage_functions}), but is more consistent with observed real world loss of load events.

Estimating the avoided loss from full winterization without using the 2021 event decreases slightly our estimates from 11.74bn\$ to 9.24bn\$. However, winterization of coal and wind is still highly profitable, and the corresponding winterized capacity for gas power is only 4GW lower than in a scenario including 2021. This indicates that even before the occurrence of the 2021 event, significant loss of load events had to be expected.

Finally, the assumed discount rate has a significant impact on results. When increasing the rate from 5\% to 7\%, the avoided loss under full winterization is reduced from 11.74bn\$ to 9.74bn\$. While winterization of coal and wind power still fully pays off under these assumptions due to low winterization cost, the profitable winterization of gas infrastructure and gas power is reduced from 13GW to 10GW.

\section*{Discussion}
We have shown that the Texas loss of load event in February 2021 was among the top three extreme events when simulating the power generation system under climate conditions of the last 71 years. In particular winterization of gas power plant and gas supply infrastructure is crucial to prevent future events. 

Our analysis indicates that events of significant loss of load had and have to be expected, which make considerable winterization efforts profitable on average. However, we have also shown that significant risk is associated with such investments, as the spread of avoided loss implies high uncertainty. Furthermore, our estimates of avoided loss are based on the price cap in the electricity market. The incentives of different actors in the power system to avoid that loss will depend on their long and short position in the market. Due to both, high risk and a somehow complex incentive structure, we see the need for regulatory intervention to enforce winterization. Apart from winterization, other flexibility measures, in particular strong demand response measures and an expansion of transmission capacities to neighbouring states, may be beneficial for the system not only during cold spells, but also to make the system more stable during the ongoing transition to a larger share of renewable energies in the power generation mix.

In a broader societal perspective the actual costs of load shedding to society, in particular during catastrophic, long-lasting events, may not be represented by the regulated price cap. This implies that benefits from winterization may even be significantly higher than our analysis indicates. Higher estimates for the value of lost load can be found in literature \cite{SocietalCosts} and others have determined the costs to society implied by the 2021 event in Texas being an magnitude of order higher than ERCOT's price cap \cite{wu2021opensource}. 

Of course, our results have to be considered in the light of a continuously evolving power system. We assume 30 years of lifetime for all installed capacities, however some capacities may soon be retired and their winterization may not be profitable therefore. As winterization of new capacity is cheaper and easier to  implement than winterization of existing one, winterization standards for installing new power plants and associated infrastructure should have high priority. The ongoing transformation of the Texan power system can therefore be considered an opportunity to ensure robustness during future cold events.


\section*{Methods}
We use a chain of statistical and simulation models to derive loss of load events, and expected avoided loss from winterization\footnote{Code related to this analysis will be available upon final publication at \url{https://github.com/inwe-boku/texas-power-outages}}. The data sets which feed the models and model interactions are shown in Figure \ref{fig:overview}. A detailed explanation of all involved models and data sets is given below.\\

\begin{figure}[!ht]
\centering
\includegraphics[scale=0.28,trim={3.5cm 15cm 0.1cm 1.1cm},clip]{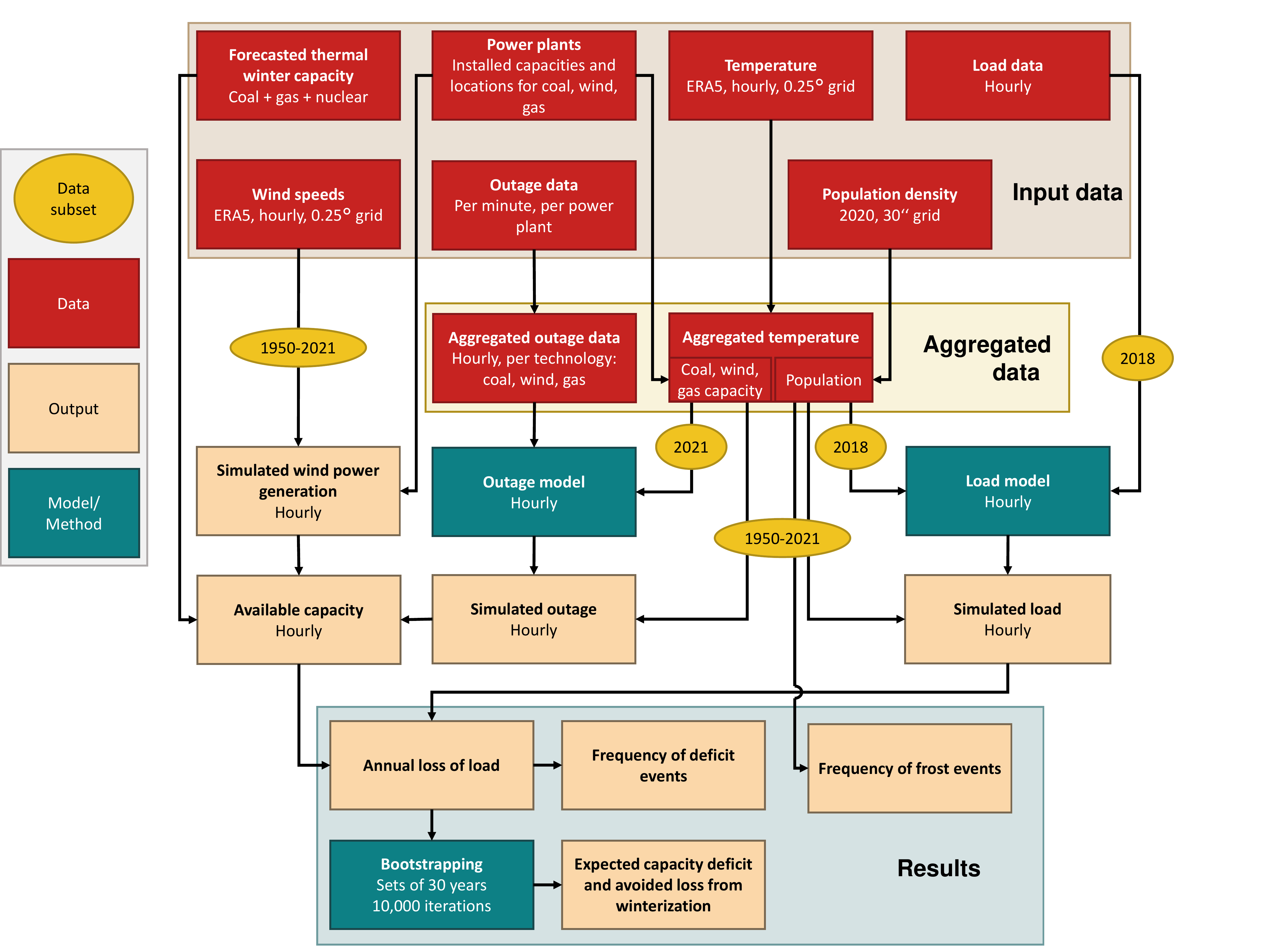}
\caption{Overview of data inputs and deficit model}
\label{fig:overview}
\end{figure}

\textbf{Estimation of temperature induced electricity deficits:} To estimate the amount of deficits in the power system, we simulate the difference between the expected, temperature dependent electricity demand and the available generation capacity.

Formally this can be described by defining the capacity deficit $d_{y,h}$ as the difference between the $\mathrm{demand}_{y,h}$ and the available generation capacity $c_{y,h}$ in year $y$ in hour of year $h$:

\begin{equation}
    d_{y,h} = \mathrm{demand}_{y,h} - c_{y,h}
    \label{eq:deficit}
\end{equation}

The total loss lof load $d^\mathrm{tot}_y$ in a year is the sum over $d_{y,h}$ for all hours where capacity deficit was larger than 0:

\begin{equation}
    d^\mathrm{tot}_{y} = \sum_{
        \substack{h,\\
        \mathrm{demand}_{y,h} \ge c_{y,h}}
    } d_{y,h}
     \label{eq:deficit_all}
\end{equation}

Demand is simulated from load in the past using a linear regression model.
Available generation capacity is obtained by reducing the total available capacity by temperature dependent outages:

\begin{equation}
    c_{y,h} = c^\mathrm{thermal} - o^\mathrm{gas}_{y,h} - o^\mathrm{coal}_{y,h} + w_{y,h}\frac{c^\mathrm{wind} -  o^\mathrm{wind}_{y,h}}{c^\mathrm{wind}}
     \label{eq:outage}
\end{equation}

where $c^\mathrm{thermal}$ is the total available thermal capacity in winter according to ERCOT, and $o^\mathrm{gas}_{y,h}$, $o^\mathrm{coal}_{y,h}$ and $o^\mathrm{wind}_{y,h}$ are the temperature dependent power plant outages, respectively. Current wind power generation $w_{y,h}$, simulated from wind speeds assuming full capacity, is reduced to the share of working wind power capacity, i.e. the difference between installed capacity $c^\mathrm{wind}$ and wind power outages $o^\mathrm{wind}_{y,h}$, divided by total installed capacity. We do not model nuclear and solar PV outages, as their overall generation capacity and their contribution to outages was low (Figure \ref{fig:load_outage_temp}). We also neglect transmission of power from neighbouring states, which is of minor magnitude.

According to ERCOT \cite{ERCOT_SARA}, 67.5GW of capacity was available during the 2021 winter, but 5GW of these may have been under maintenance. We therefore assume a value of 62GW for $c^\mathrm{thermal}$ to match our simulation with available capacity during the event, as indicated by served load.\\

\label{section:demand_prediction}
\textbf{Demand prediction:} We predict $\mathrm{demand}_{y,h}$ from temperatures using a regression model (equation \eqref{eq:load}). For each year $y$ in the period 2018--2020 we estimate a model for demand in winter (December--February, the three coldest months on average), taking into account seasonality (sine and cosine terms depending on the hour of the year $h$), and the temperature $t_{y,h}$. Furthermore, we include dummy variables for the hour of day $\delta^\mathrm{hod}_{i,y,h}$, the weekdays $\delta^\mathrm{dow}_{i,y,h}$, and for holidays $\delta^\mathrm{hold}_{y,h}$. We include $t_{y,h}^4$ into the equation as with falling temperatures, load shows a highly non-linear increase. We tested a quadratic, a cubic and a quartic term and the regression with the quartic term showed the highest fit.

\begin{equation}
\label{eq:load}
\begin{split}
    \mathrm{demand_{y,h}} = {} & \beta_0 + 
    \sum_{i=1}^{7} \beta_i \delta_{i,y,h}^\mathrm{dow} +
    \sum_{j=8}^{31} \beta_j \delta_{j,y,h}^\mathrm{hod} +
    \beta_{32} \delta^\mathrm{hold}_{y,h} + \\
    & \beta_{33} \sin\left(\frac{2\pi h}{8760}\right) +
    \beta_{34} \cos\left(\frac{2\pi h}{8760}\right) +
    \beta_{35} t_{y,h} + \beta_{36} t_{y,h}^4
\end{split}
\end{equation}

\begin{figure}[!ht]
\centering
\includegraphics[scale=0.47,trim={0.1cm 0.1cm 0.1cm 0.1cm},clip]{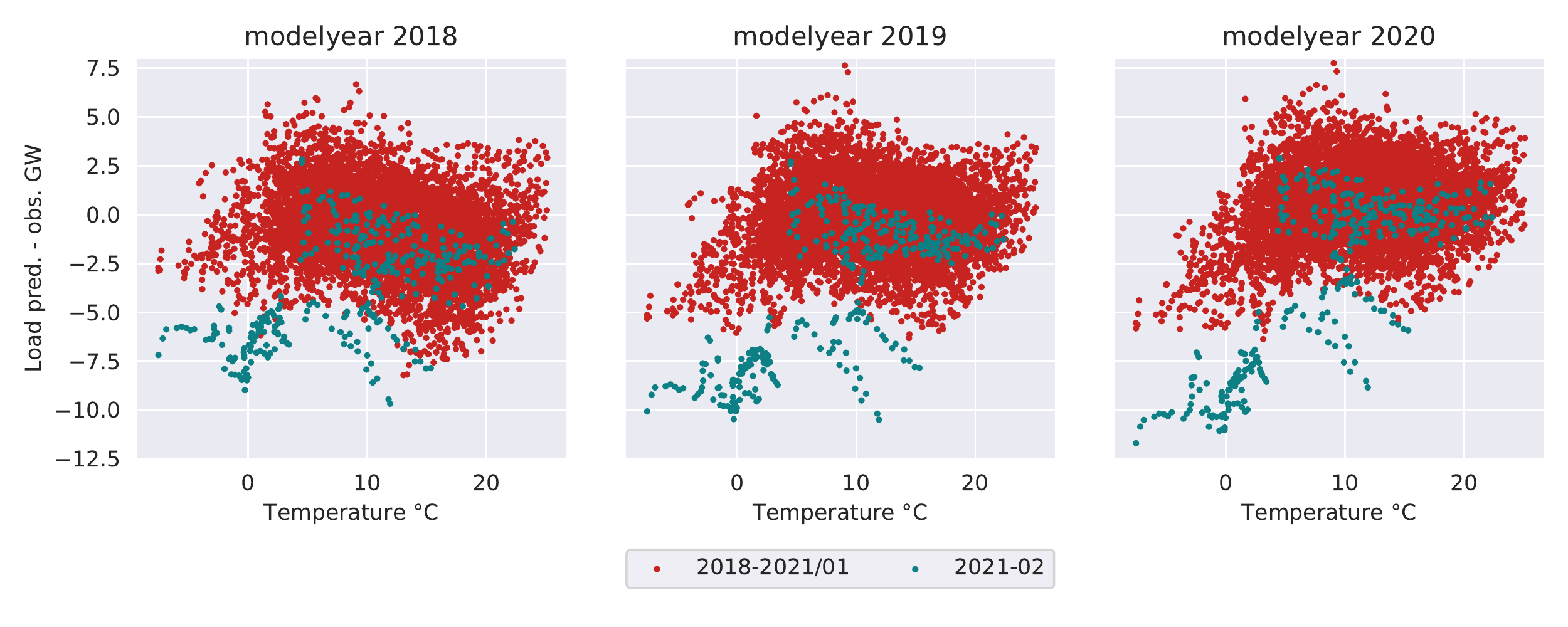}
\caption{Difference between predicted and observed load applying the regression model estimated from 2018, 2019 and 2020 data to the period 2018--2021/01 and the days before the blackout (2021/Feb/01--2021/Feb/14)}
\label{fig:load_fit_modelyear}
\end{figure}

We applied the models estimated with data from the years 2018, 2019 and 2020, respectively, to 2021. The total load differs by a maximum of 2\% in the three years. When predicting with the models for January 2021, the 2020 model shows the highest fit (RMSE 1.75GW, $R^2$ 0.87, 2019: RMSE 2.31, $R^2$ 0.77, 2020: RMSE 2.96, $R^2$ 0.62). However, using the 2019 and 2020 model substantially and consistently underestimates loads at very low temperatures. This is probably due to relatively warm winters in these years. The 2018 model performs better under these conditions, although load estimates are still lower than observed at the lower temperature end (Figure \ref{fig:load_fit_modelyear}). In particular in 2021, all three models underestimate the load in the hours before the blackouts, although temperatures weren't record low during this time. This warrants further research.  \\

\textbf{Temperature dependent generator outages:} 
The power plant outages in Texas were caused by a systemic failure, as a substantial amount of capacity tripped in a short period of time \cite{OutageCauses}. Most of the capacity failed due to weather events, but the lack of fuel supply was also relevant. Additionally, non- weather related equipment failures and, to a smaller amount, frequency problems in the grid caused problems at generation units. This implies that temperature at single power plants alone cannot fully explain outages. We derived infrastructure fragility curves \cite{Dunn2018} from our data and they show that 25\% of gas power plants, about 30\% of wind power plants, and 50\% of coal power plants failed when temperatures were above 0°C and in many cases have not been below 0°C in the days before failure. Other weather conditions besides freezing temperatures may be partly responsible, but at least wind conditions at the time of failure were not particularly extreme. Besides freezing temperatures, which caused failure of power plant components and of gas production and gas distribution, there are also systemic reasons for failure therefore.

We do not use infrastructure fragility curves for modeling outages, but estimate threshold temperatures and outage capacities during catastrophic failure by technology: we derive the temperature, at which the largest increase in outages occurred, from the 2021 outage data. Furthermore, we estimate an outage level in terms of tripping capacity. Finally, we also define a constant recovery rate, which describes how outages decreased after the recovery temperature is reached. This is an aggregate approach and will omit smaller outages, but it accounts better for the inter-dependency of failures. A detailed outline of how we derived the outage parameters from the 2021 data in Texas is given in section \ref{subsection:outage_functions}. The resulting outage models are subsequently applied to the 71 years of climate data to obtain capacity availability during this period. In a sensitivity analysis, we have also tested how loss of load changes if a similar outage model on the level of individual plants is used instead. \\

\textbf{Estimating avoid loss of winterization:} We use the price cap of 9\,000\$/MWh regulated by ERCOT \cite{Pallone2021} as estimate of the value of lost load. We first derive the loss of load $d^\mathrm{tot}_{y,p,w}$ for different scenarios, where a capacity $w$ of power plants of technology $p$ (gas, coal or wind in North or South Texas) are winterized. This is done similar to equation (\ref{eq:deficit_all}) but the maximum outage level for a technology in the outage functions $o^p_{y,h}$ used in equation (\ref{eq:outage}) is reduced by $w$. We estimate total loss of load $d^\mathrm{tot}_{y,p,w}$ using temperature data of all years in the period 1950--2021. We combine temperature dependent demand estimates, simulated wind power generation for the same years from ERA5 reanalysis wind speeds (for further details see \cite{gruber2020global}), and the temperature dependent power plant outage functions for that purpose. The 71 different weather years therefore serve as a set of different weather realisations where we bootstrap from. We assume that the power system is equivalent to today's system. 
For each bootstrapping iteration $b$, we sample 30 years of lost loads $d_{i,b,p,w}^\mathrm{boot}$ for $i=1,\ldots,30$ from the set of $d^\mathrm{tot}_{y,p,w}$, $y=1950,\ldots,2021$. This is done for 10\,000 iterations, $b=1,\ldots,10\,000$. 
Subsequently, total economic loss $l_{b,p,w}$ over a typical investment period of 30 years is calculated, as shown in equation~\eqref{eq:bootstrap}, $r=0.05$ being the discount rate:
\begin{equation}
\label{eq:bootstrap}
\begin{split}
    l_{b,p,w} = 9\,000 \sum_{i=1}^{30} d^\mathrm{boot}_{i,b,p,w}(1+r)^{-i}
\end{split}
\end{equation}

This results in 10\,000 different samples $l_{b,p,w}$, allowing to derive a distribution of economic losses under different scenarios of winterization. To calculate avoided loss from winterization by technology $p$, we derive $l^\mathrm{avoid}_{b,p,w}$ in the following way:
\begin{equation}
\label{eq:revenues}
\begin{split}
    l^\mathrm{avoid}_{b,p,w} = l_{b,p,0} - l_{b,p,w}
\end{split}
\end{equation}
We eventually take the first differences of $l^\mathrm{avoid}_{b,p,w}$ to derive the marginal avoided loss from winterization. \\

\textbf{Frequency analysis of frost and power deficit events}: The frequency analysis of extreme events follows well-established methods of hydrological drought analysis \cite{Laaha2017}, where deficit events are defined as periods when the variable of interest is below a certain threshold. Here, we use two different threshold concepts. First, we analyse temperature, and define a constant threshold of 0°C to define deficit events in analogy to drought events in drought statistics. Deficit events in the power system, resulting from low temperatures, are simply defined for periods when the capacity deficit $>$ 0GW (equation \eqref{eq:deficit}). In each case, the result is a derived deficit time series, which is further investigated using Yevjevich’s theory of runs \cite{yevjevich1967objective}.
During a frost period, minor thaw episodes or other disturbances may split an event in several smaller events. As a remedy, pooling procedures have been recommended \cite{Tallaksen2004}. In this study, an inter-event time criterion of 1 day is used to define the deficit event series. In case that multiple events occur in a year the event with the absolutely largest accumulated deficit is used for further analysis.
The series so derived are characterized by three deficit characteristics: duration (measured in hours), intensity (minimum temperature / maximum power deficit), and severity (aggregated frost sum / power deficit over the event), each of which constitutes an annual extreme value series. These are further analyzed using extreme value statistics to determine the return period of each frost and capacity deficit characteristic according to natural hazard management standards. Our analysis was conducted using the R-software package lfstat \cite{lfstat}, which provides a collection of state-of-the-art methods that are fully described in the World Meteorological Organization's manual on low flow estimation and prediction \cite{Gustard2008}. The resulting  winter power deficit events are shown in Figure \ref{fig:deficit_events}, their extreme value statistics are shown in Figure \ref{fig:extreme_statistics}.\\

\textbf{Data:} Temperature at 2m above ground is taken from the ERA5 reanalysis \cite{ERA5}. We weight temperature over Texas by population density \cite{GPW} to derive a temperature index for modeling electricity demand.
For estimating outages in the power system due to low temperatures, we also weight temperature by capacities of wind \cite{USWTDB}, coal and natural gas power plants \cite{powerplants}, which were the power generation technologies most affected by failure during the extreme temperature event of February 2021. For wind power plants, we split into a North and South region (along the latitude of 30°), since temperatures at wind parks in the North and South of Texas differ substantially.
Since the failure of the power system might be related to infrastructure at gas fields supplying these power plants \cite{TXwinterize}, we also determine a gas field weighted temperature index, using the distribution of natural gas production by county \cite{gasfields} to complement our analysis. 
Load data used for demand prediction were retrieved from ERCOT \cite{ERCOTload} for the period 2004--2021/02. Since the focus of this study is on cold events in winter, only winter load data (Dec--Feb) is used.
Outage data is provided in the period since 10\textsuperscript{th} of February 2021 by ERCOT \cite{outages} and is aggregated by power generation technology for the analysis.

\section*{Acknowledgement}
We gratefully acknowledge support from the European Research Council (“reFUEL” ERC2017-STG 758149). We are grateful to Edgar Virgüez who provided spatial locations of power plant outages and discussed early results with us.

\printbibliography

\newpage
\appendix
\renewcommand\thefigure{\thesection.\arabic{figure}}
\setcounter{figure}{0}
\renewcommand\thetable{\thesection.\arabic{table}}
\setcounter{table}{0}
\section{Appendix}

\subsection{Load and temperature extremes}

\begin{figure}[!h]
\centering
\includegraphics[scale=0.68,trim={0.2cm 0.59cm 0.2cm 0.4cm},clip]{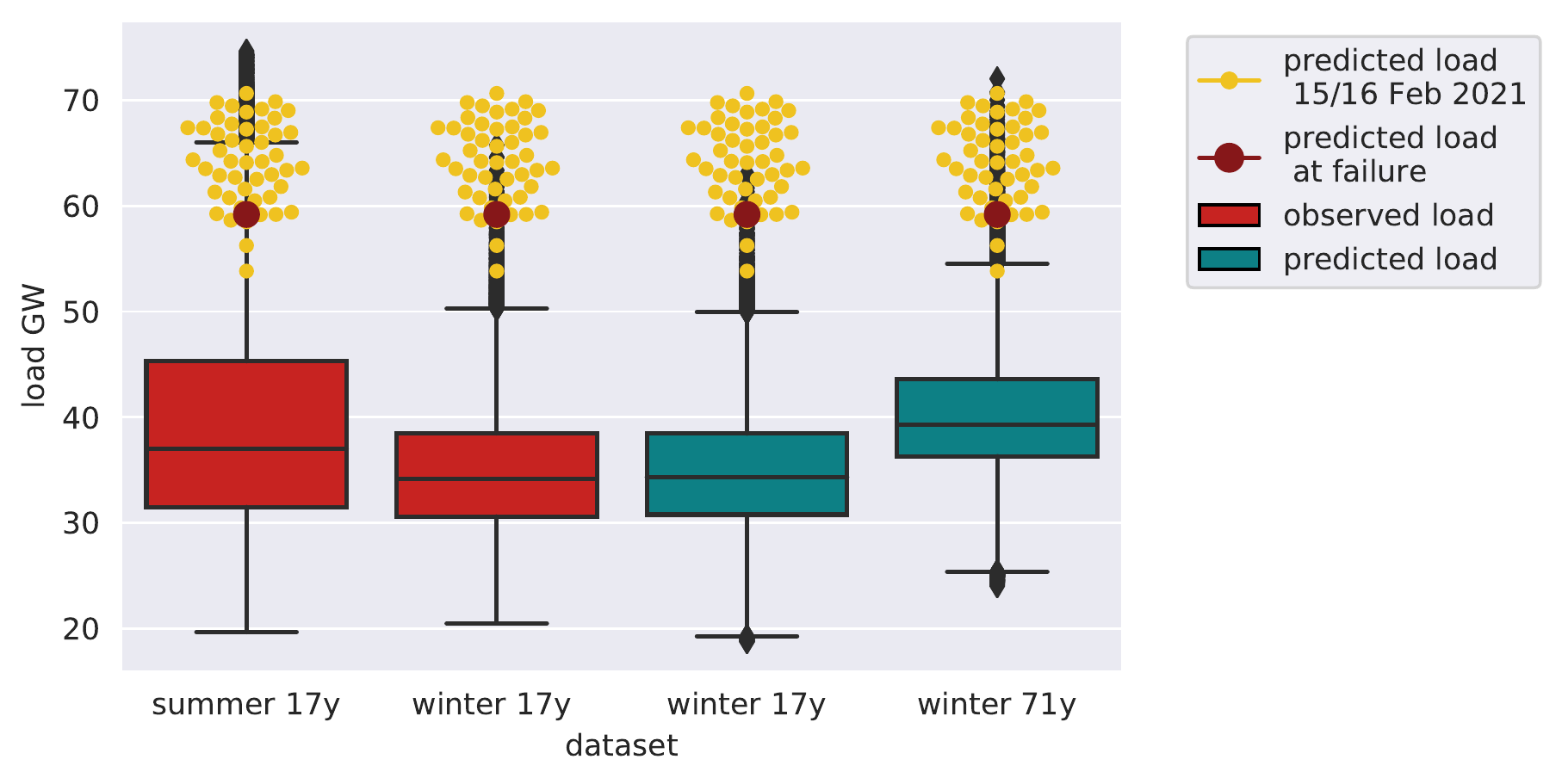}
\caption{Observed loads for 17 years (2004--2021/01) for winter (Dec--Feb) and summer (Mar--Nov)) periods, predicted winter loads for 17 years (2004--2021/01) and 71 years (1950--2021/01) and extreme loads on February 15\textsuperscript{th} and 16\textsuperscript{th} 2021}
\label{fig:load_pred_obs_vs_Feb2021}
\end{figure}

\begin{figure}[!h]
\centering
\includegraphics[scale=0.68,trim={0.1cm 0.1cm 0.1cm 0.1cm},clip]{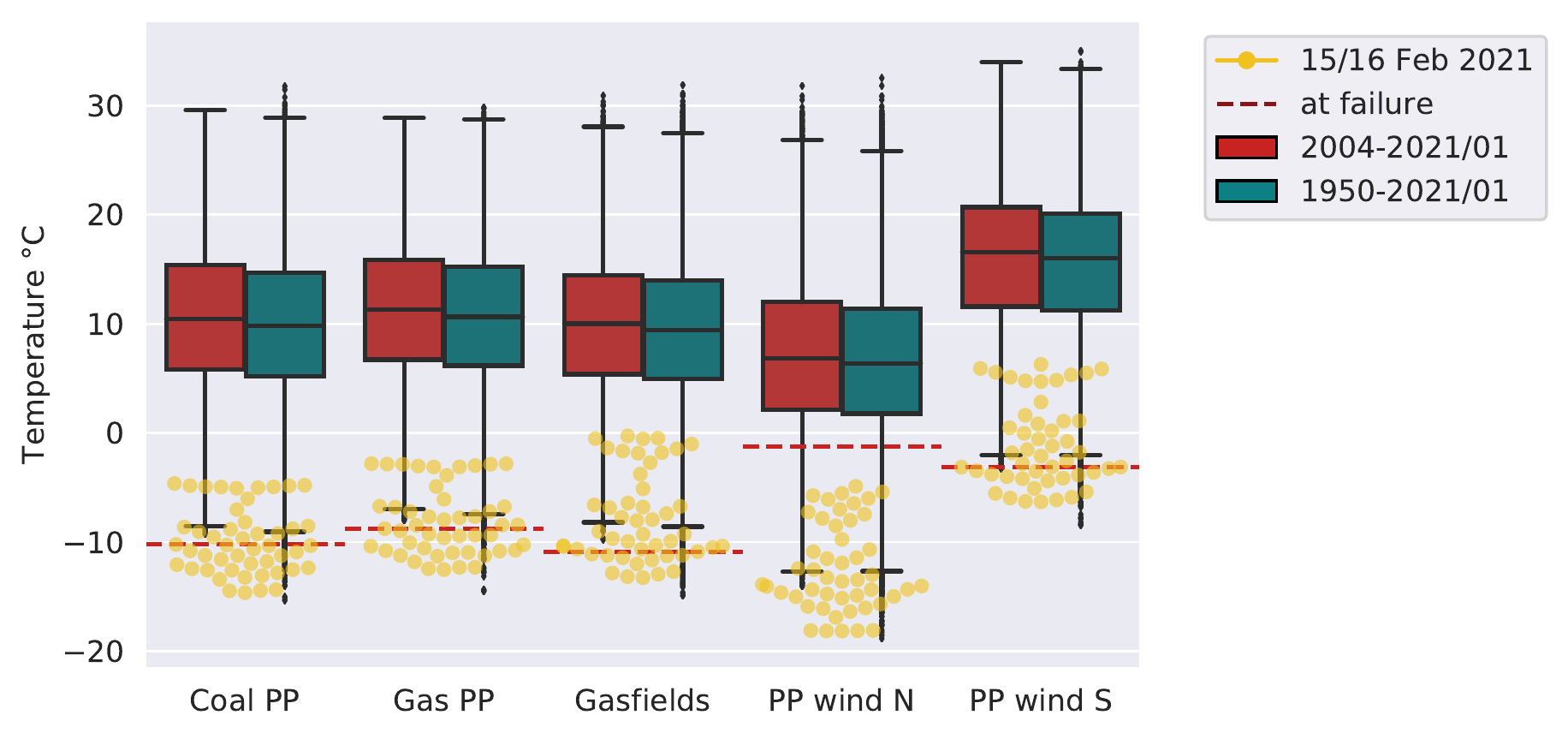}
\caption{Winter temperatures weighted by natural gas and power plant capacity, gas field production and wind power plant capacity in North and South Texas compared to temperatures during the 15/16 Feb 2021 event}
\label{fig:temp_gasfields_powerplants}
\end{figure}

\subsection{Extreme value statistics of temperature and power deficits}

\begin{figure}[!h]
\centering
\includegraphics{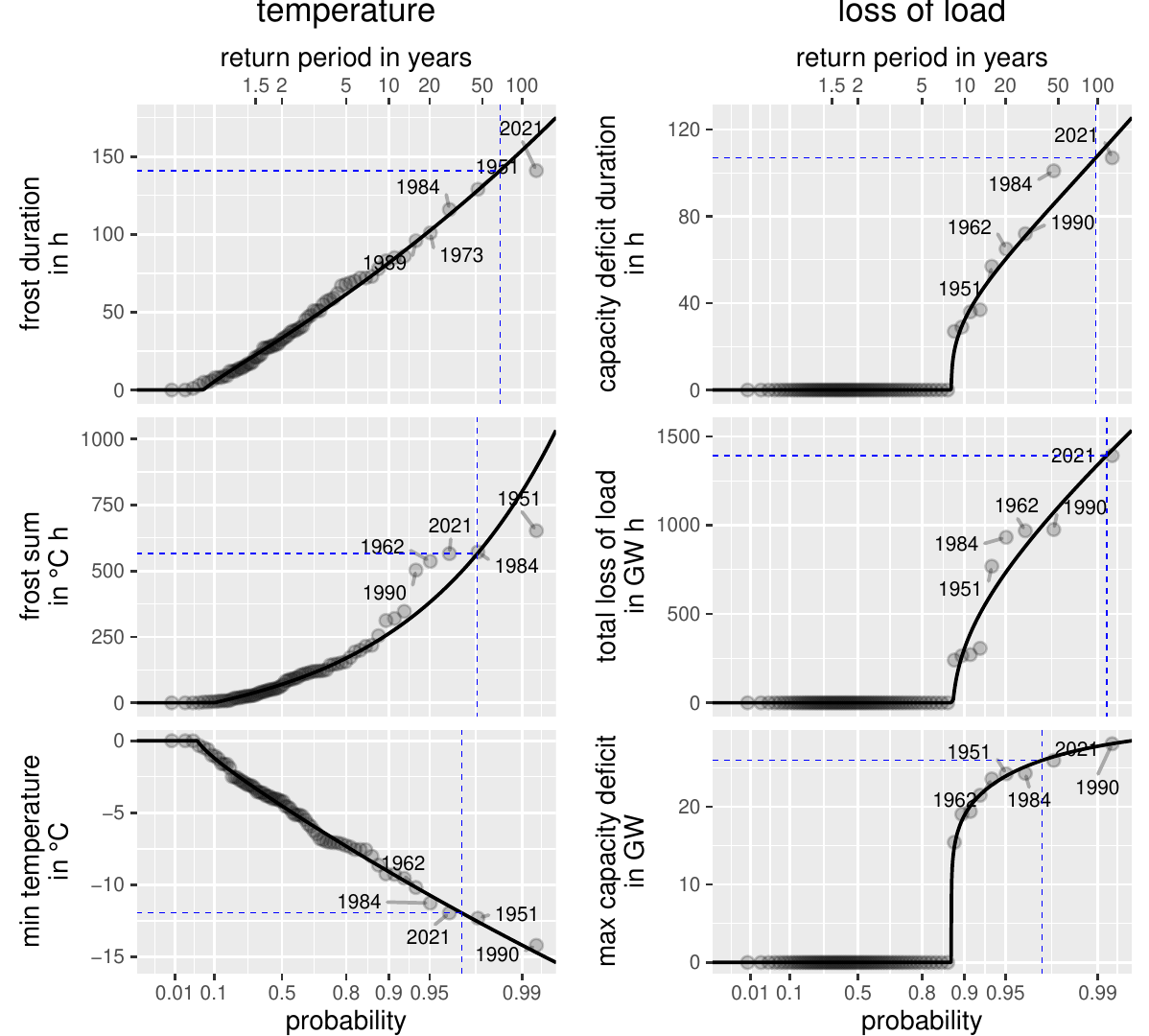}
\caption{Extreme value statistics of population weighted temperature (left panels) and predicted capacity deficit (right panels) of frost events in Texas from seven decades of of climate data. Shown are the empirical (circles) and theoretical (lines) quantile functions of the three deficit characteristics, duration (upper), severity (centre) and intensity (lower). The return period (inferred from a GEV distribution) and magnitude of the 2021 event are annotated in blue}
\label{fig:extreme_statistics}
\end{figure}

\newpage
\subsection{Visualization of load prediction model}

\begin{figure}[!ht]
\centering
\includegraphics[scale=0.8,trim={0.5cm 0.1cm 0.1cm 0.1cm},clip]{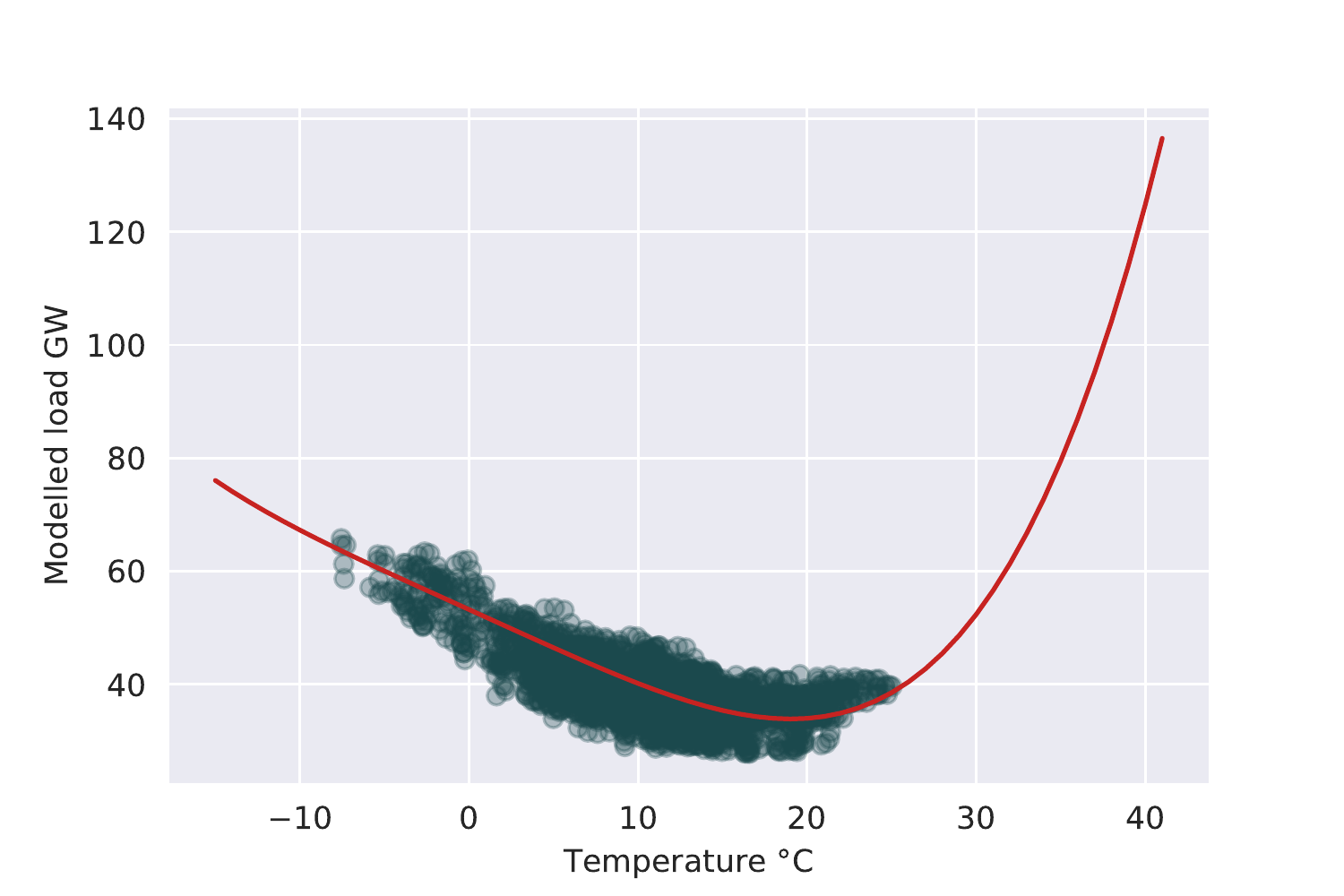}
\caption{Predicted load from 2018 data with constant time parameters for the range of Texas temperatures and load data for the winter period of 2018}
\label{fig:model_range}
\end{figure}

\subsection{Approximation of outage functions}

\label{subsection:outage_functions}
\begin{figure}[!ht]
\centering
\includegraphics[width=\textwidth]{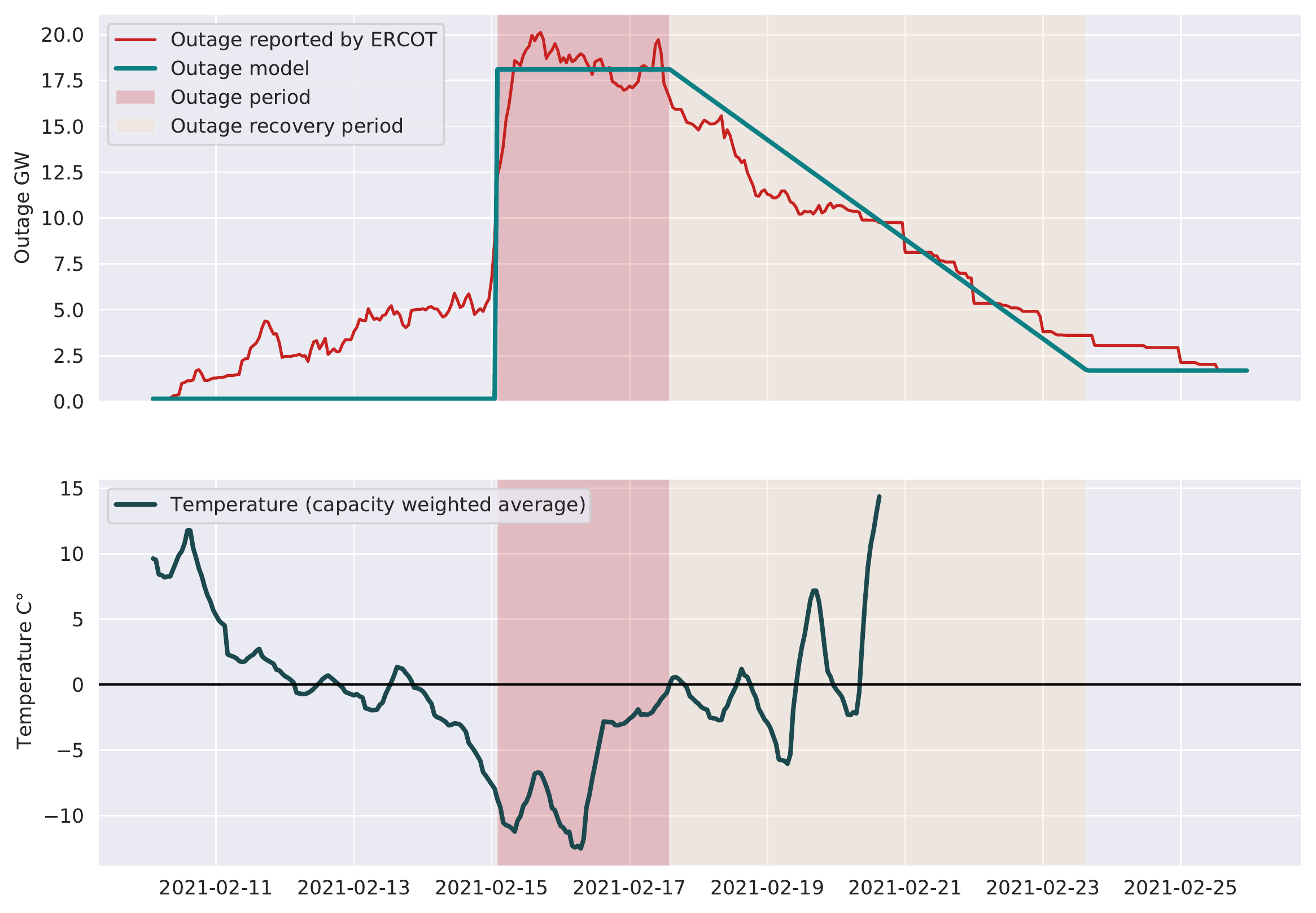}
\caption{Gas outages and temperature at gas power plants and approximated outage model}
\label{fig:thresh_gas}
\end{figure}

\begin{figure}[!ht]
\centering
\includegraphics[width=\textwidth]{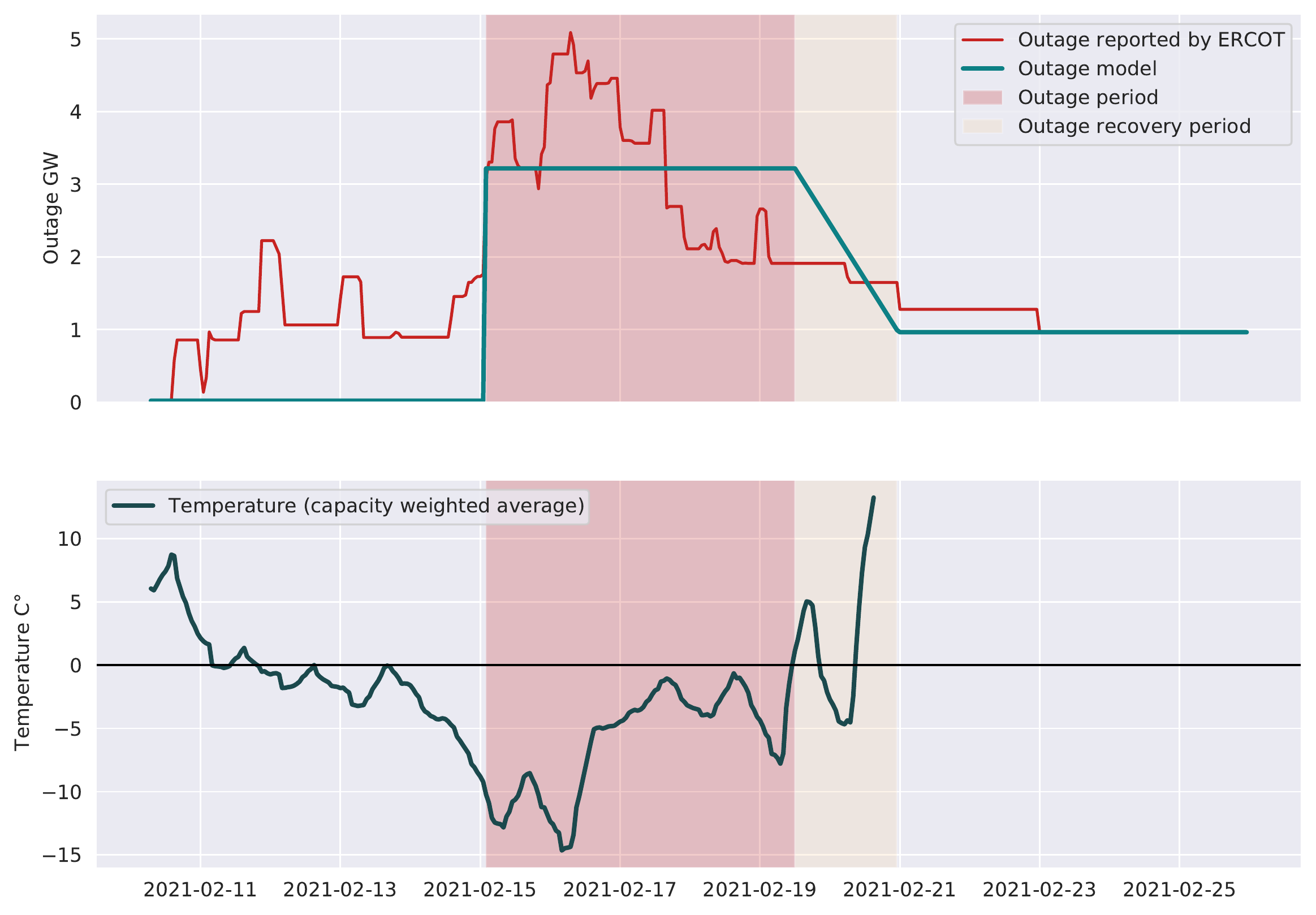}
\caption{Coal outages and temperature at coal power plants and approximated outage model}
\label{fig:thresh_coal}
\end{figure}

\begin{figure}[!ht]
\centering
\includegraphics[width=\textwidth]{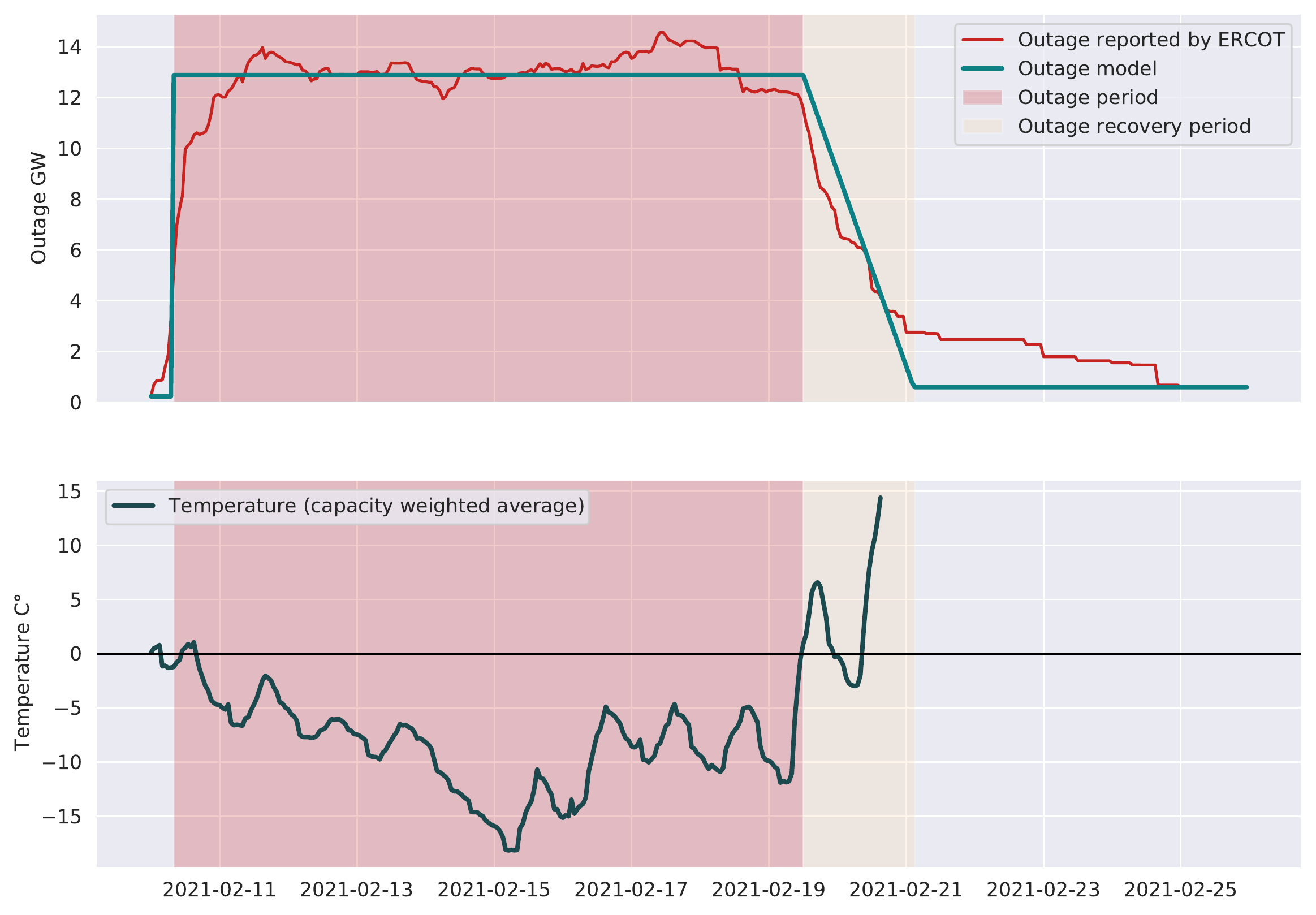}
\caption{Wind outages and temperature at wind power plants in Northern Texas and approximated outage model}
\label{fig:thresh_wind_north}
\end{figure}

\begin{figure}[!ht]
\centering
\includegraphics[width=\textwidth]{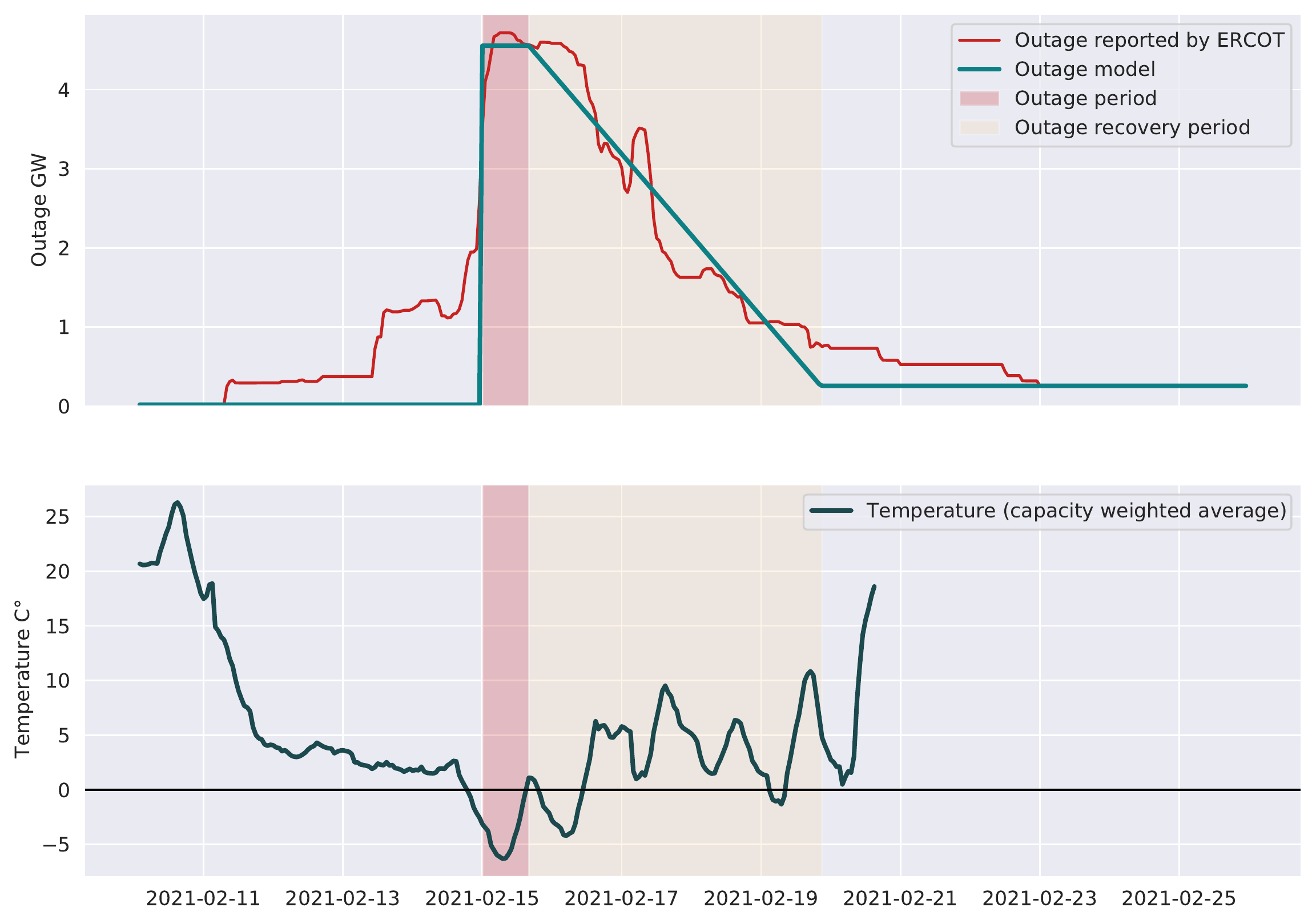}
\caption{Wind outages and temperature at wind power plants in Southern Texas and approximated outage model}
\label{fig:thresh_wind_south}
\end{figure}

We model the outages of the sum of capacities by technological group, defining an outage model with 4 segments: (1) constant outage level before major critical failure, (2) constant outage level during critical failure, (3) declining outage in recovery period, and (4) constant outage level after recovery period (see Figures \ref{fig:thresh_gas} to \ref{fig:thresh_wind_south}).
(1) The level of the first segment is defined by the first data point in the outage time series. The first segment ends when the single largest increase in outage within one hour occurs (i.e. the maximum of the first differences of the outage time series). The average power plant capacity weighted temperature at that point is defined as outage temperature. (2) The second segments starts at that point and ends when the temperature increases above the recovery temperature, which is set to 0°C consistently for all technologies. However, we neglect any hours with temperatures above the recovery temperature within the first 10 hours after the start of the second segment. The level of the plateau is derived by ensuring that the area below the real outage curve is the same as the area below our modelled outage. (3) and (4) We extract the outage data from the point of recovery to the end of the timeseries on the 25\textsuperscript{th} of January. We then fit a model to the data, which minimizes least squares. It contains two segments: one falling recovery segment, and one constant segment at the end of the timeseries - the level of that constant segment is defined as the average of the last ten points in the time series. For segment (3), we can derive a slope after fitting, which is used as parameter to simulate recovery. 
These models are applied to the whole 71 year long time series of temperature data, removing the constant outages at the beginning (1) and the end (4). An outage starts in the model, whenever the power plant capacity weighted temperature falls below the outage threshold. The full outage lasts until the temperature increases above the recovery temperature, but at least 10 hours. From that moment on, a linear, falling outage is assumed, until the outages is reduced to 0GW. 

\clearpage

\subsection{Trends in extreme events}
\label{section:trends_extreme}
\begin{figure}[!ht]
\centering
\includegraphics{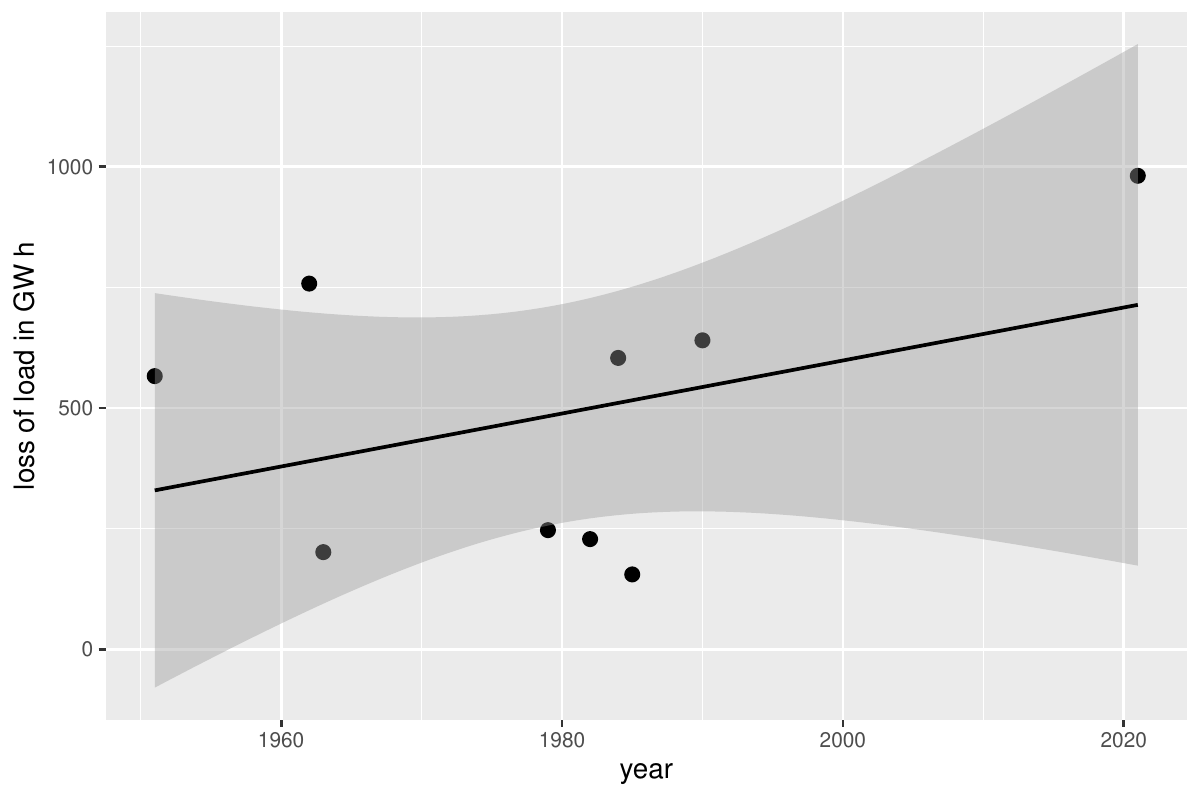}
\caption{Loss of load events in the period 1950--2021 with non-significant trend line (p-value=0.31)}
\label{fig:trend-9events}
\end{figure}

\subsection{Modeling climate change trends in temperature}
\label{section:avg_tmep_trend}
\begin{figure}[!h]
\centering
\includegraphics[scale=0.7,trim={0.1cm 0.6cm 0.1cm 0.1cm},clip]{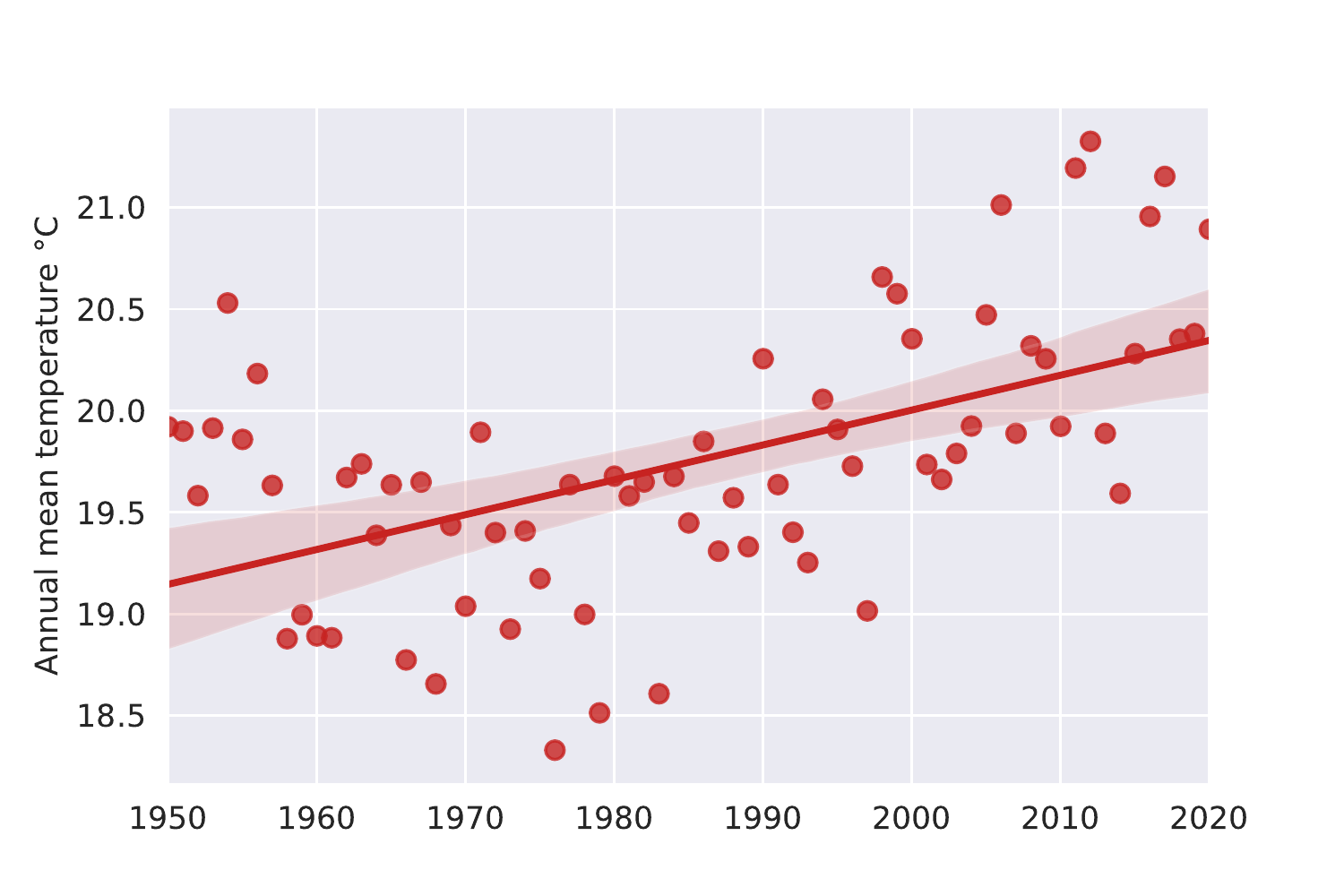}
\caption{Trends in population weighted annual mean temperatures}
\label{fig:avg_temp_trend}
\end{figure}

Figure \ref{fig:avg_temp_trend} shows mean yearly population weighted temperatures over the past 71 years. Regression indicates a clear trend towards higher average temperatures with an annual increase of 0.017°C. 

In order to introduce an artificial linear trend to past temperatures, we increased all past temperatures by the trend observed in average temperature since 1950 assuming the linear trend of average temperatures would also translate to extreme temperatures, although there is no evidence of this phenomenon.
Using the estimated average temperature trend of 0.017°C per year in the period 1950--2020, temperatures in the time series are updated in the following way:
\begin{equation}
t^\mathrm{trend}_{y,h}=t_{y,h} + 0.017(y_\mathrm{ref} - y)
\end{equation}
We calculated two scenarios, using 2021 and 2050 as $y_\mathrm{ref}$. These are the boundaries for our analysis, when considering a 30 years investment period.

\subsection{Trends in extreme temperature}
The predicted outage relies on the stationarity assumption for the temperature time series, in particular on the assumption that there is no trend in temperatures.
We use past climate data to simulate outages. Due to climate change, we however observe an increase in average temperatures in Texas. Our estimates may therefore be biased. Still, extreme events in the power system, in our model, occur at extremely low temperatures for the Texan context. These extreme cold events do not necessarily follow the trend in the average increase of temperatures \cite{extremeTrend}. Our model uses a minimum threshold of -10.2 °C for coal, -8.8 °C for gas, -10.9 for gasfields, -1.2 °C for wind in Northern Texas and -3.1 °C for wind outages in Southern Texas. Therefore, we specifically need to examine stationarity of annual minimum temperatures below this threshold. By conducting a robust trend analysis for annual temperature minima below the threshold for different temperature thresholds, we find that extreme frost events in Texas only show a trend if the temperature threshold is set to -1°C or above, i.e. very high. For temperature thresholds below -1°C, no trend can be confirmed (see Table \ref{tab:trend_extreme_temp}). Assuming a trend in a reduction of extreme cold events below -1°C can therefore not be confirmed by the 71 years of temperature data available to us. \\ 
\label{section:ext_temp_trends}
\begin{table}[h!]
\caption{Trends in cold temperature with different temperature thresholds}
\begin{tabular}{p{1.7cm}|p{1cm}|p{1.5cm}|p{1.5cm}|p{1.5cm}}
\hline
temperature threshold & events & slope yearly & slope 71 years & p-value \\ \hline
0 & 71 & 0.037 & 2.65 & 0.029\\
-1 & 70 & 0.037 & 2.61 & 0.033\\
-2 & 63 & 0.020 & 1.43 & 0.256\\
-3 & 55 & 0.004 & 0.30 & 0.816\\
-4 & 38 & -0.005 & -0.38 & 0.810\\
-5 & 30 & -0.002 & -0.11 & 0.948\\
-6 & 25 & 0.013 & 0.95 & 0.547\\
-7 & 20 & 0.018 & 1.27 & 0.474\\
-8 & 10 & -0.028 & -1.96 & 0.316\\
-9 & 8 & -0.017 & -1.22 & 0.422\\
-10 & 5 & -0.013 & -0.89 & 0.491
\end{tabular}
\label{tab:trend_extreme_temp}
\end{table}

\subsection{Calculation of winterization costs}
\label{section:winterization_costs}
There is very limited information on winterization costs available from media reports. 
For gas wells, costs of 50,000\$ for winterization are reported \cite{TXwinterize}. Winterizing all 123,000 gas wells in Texas \cite{gasWells} would therefore yield a total cost of 6.15bn\$. 18GW of gas power capacity failed during the 2021 event according to our simulation.  Conservatively assuming that winterization costs of gas fields can be split according to failed gas power capacity, winterization costs for gas fields of 342 Mio\$/GW of gas power plant capacity can be derived. This is equivalent to around 250GWh of pipe storage for methane, which could be installed on the site of gas power plants to secure supply under cold conditions as alternative \cite{Kruck2013Overview}. Assuming costs of gas power plant winterization to be 10\% of investment costs, total winterization costs result in 453 Mio\$/GW of gas power capacity at an investment costs for gas power plants of 1.12bn\$/GW \cite{DanishEnergyAgency2016Technology}.\\
For coal and wind power plants, no infrastructure has to be winterized. Therefore, winterization costs will be significantly lower than for gas. Winterization of wind turbines is about 5\% \cite{wind_winterize} of investment costs. Assuming investment costs of to 1.3bn\$/GW \cite{ARMIJO20201541}, this yields 65 Mio\$/GW of wind power capacity as estimate of winterization costs for wind turbines.\\
For coal power plants, we did not find any estimate, and assume 10\% of investment costs. At investment costs for coal power plants of 2.24 bn\$/GW \cite{DanishEnergyAgency2016Technology}, winterization costs of 224 Mio\$/GW are obtained.
\end{document}